\documentclass[11pt]{article}

\usepackage{epsfig,amssymb,euscript}
\usepackage{amsmath,amscd}
\usepackage{amsfonts}

\numberwithin{equation}{section}
\newcommand{\be}{\begin{eqnarray}}
\newcommand{\ee}{\end{eqnarray}}
\newcommand{\bea}{\begin{eqnarray}}
\newcommand{\eea}{\end{eqnarray}}
\newcommand{\ba}{\begin{array}}
\newcommand{\ea}{\end{array}}

\newcommand{\bR}{\mathbb{R}}
\newcommand{\bC}{\mathbb{C}}

\def\appendix#1{\addtocounter{section}{1}\setcounter{equation}{0}
\renewcommand{\thesection}{\Alph{section}}
\section*{Appendix \thesection\protect\indent \parbox[t]{11.15cm}{#1}}
\addcontentsline{toc}{section}{Appendix \thesection\ \ \ #1}}

\begin{document}

\begin{titlepage}

\vfill

\begin{flushright}
\end{flushright}

\vfill

\begin{center}
\baselineskip=16pt
{\Large\bf Para-Complex Geometry and Gravitational Instantons}
\vskip 2cm
J. B.  Gutowski$^1$ and W. A. Sabra$^2$
\\
\vskip .6cm
\begin{small}
$^1$\textit{  Department of Mathematics, King's College London
\\
Strand, London WC2R 2LS, UK.\\
E-mail: jan.gutowski@kcl.ac.uk}
\end{small}\\*[.6cm]
\begin{small}
$^2$\textit{Centre for Advanced Mathematical Sciences and
Physics Department, \\
American University of Beirut, Lebanon \\
E-mail: ws00@aub.edu.lb}
\end{small}\\*[.6cm]
\end{center}
\vfill

\begin{center}
\textbf{Abstract}
\end{center}

\begin{quote}
We give a complete classification of supersymmetric gravitational instantons in
Euclidean N=2 supergravity coupled to vector multiplets.
An interesting class of solutions is found which corresponds to the Euclidean
analogue of stationary black hole solutions of N=2 supergravity
theories.
\end{quote}

\vfill

\end{titlepage}

\section{Introduction}

Instantons are of particular importance in theoretical
physics and mathematics. For example, instantons are an essential ingredient
in the non-perturbative analysis of non-Abelian gauge theories and quantum
mechanical systems \cite{instanton-lectures}. Moreover, the existence of
spin-1/2 zero-mode of the instanton is linked to the Atiyah-Singer index
theorem \cite{atiyah}, a fact reflecting the intimate relation of
non-Abelian gauge theory to the field of fibre bundles and differential
geometry. An important example
of Yang-Mills instanton solutions are those given in \cite{polyakov}. In
finding the instanton solutions of \cite{polyakov}, the self-duality (or
anti-self-duality) is imposed on the Yang-Mills field strength, this leads
to the fact that the Bianchi identity implies the Yang-Mills field
equations. This considerably simplifies  finding solutions, as
instead of solving second order differential equations, one solves
the Bianchi identities containing only first derivatives of the vector
potential. Gravitational instantons are in general defined as non-singular
complete solutions to the Euclidean Einstein equations of motion. Notable
early examples of gravitational instantons are the Eguchi-Hanson instantons 
\cite{eguchi}. which are the first examples of the family of the
Gibbons-Hawking instanton solutions \cite{Hawking}.

In finding gravitational instantons \cite{eguchi, Hawking} and in analogy
with the Yang-Mills case, the spin connection one-form is assumed to be
self-dual (or anti-self-dual), which leads to a self-dual curvature
two-form. This property together with the cyclic identity ensures that
Einstein's equations of motion are satisfied. The equations coming from the
self-duality of the spin connection are simpler as they contain only first
derivatives of the spacetime metric.

In recent years, a good deal of work has been done on the classification of
solutions preserving fractions of supersymmetry in supergravity theories in
various dimensions. It is clear that the quest of finding solutions
admitting some supersymmetry is easier as one in these cases is simply
dealing with first order Killing spinors differential equations rather than
Einstein's equations of motion. Following the results of \cite{hullgib}, a
systematic classification for all metrics admitting Killing spinors in $D=4$
Einstein-Maxwell theory, was performed in \cite{Tod}. The solutions with
time-like Killing spinors turn out to be the  IWP (Isreal-Wilson-Perj\'es) solutions \cite{IWP} whose
static limit is given by the the Majumdar-Papapetrou solutions \cite{mp}. It was shown by
Hartle and Hawking that all the non-static solutions suffered from naked
singularities \cite{naked, anaked}. Using the two-component spinor calculus 
\cite{penrose}, the instanton analogue of the IWP\ metric was constructed in 
\cite{DH07}. These solutions were also recovered in the complete
classification of instanton solutions admitting Killing spinors using
spinorial geometry techniques \cite{instantons}. Spinorial geometry, partly
based on \cite{lawson, wang, harvey}, was first used in \cite{first} and
has also been a very powerful tool in the classification of solutions in
lower dimensions (see for example \cite{recentlower}) and in the
classification of supersymmetric solutions of Euclidean $N=4$ \ super
Yang-Mills theory \cite{dietmar1}.

Sometime ago general stationary solutions of $N=2$ supergravity action
coupled to $N=2$ matter multiplets were found in \cite{BLS}. These can be
thought of as generalizations of the IWP\ solutions of Einstein-Maxwell
theory to include more gauge and scalar fields. The symplectic formulation
of the underlying special geometry played an important role in the
construction of these solutions. The stationary solutions found are
generalization of the double-extreme and static black hole solutions found
in \cite{sab}. It was also shown in \cite{ortin} that the solutions of 
\cite{BLS} are the unique half-supersymmetric solutions with time-like
Killing vector. The $N=2$ solutions are covariantly formulated in terms of
the underlying special geometry. The solution is defined in terms of the
symplectic sections satisfying the so-called stabilization equations.

In the present work we extend the construction of \cite{instantons} to $N=2$
Euclidean supergravities with gauge and scalar fields. A class of these
theories were recently derived in \cite{mohaupt3} as a reduction of the
five-dimensional $N=2$ supergravity theories coupled to vector multiplets 
\cite{GST} on a time-like circle. The paper is organized as follows. In the
next section, we will collect some formulae and expressions of $N=2$
supergravity which will be important for the following discussion. Section
three contains a derivation of the gravitational instantons using spinorial
geometry method. The solutions found are the Euclidean analogues of the
stationary black hole solutions of \cite{BLS}. Section four contains a
summary and some future directions. We include an Appendix containing a
linear system of equations obtained from the Killing spinor equations.

\section{Special Geometry}

In this section we review some of the structure and equations of the
original theory of special geometry when formulated in $(1,3)$ signature. We
then briefly discuss the modifications one introduces for the Euclidean 
$(0,4)$ signature. For further details on the subject the reader is referred
to \cite{speone}. The bosonic Lagrangian of the four-dimensional $N=2$
supergravity theory coupled to vector multiplets can be written as

\begin{equation}
\mathbf{e}^{-1}\mathcal{L}=\frac{1}{2}R-g_{A\bar{B}}\partial _{\mu
}z^{A}\partial ^{\mu }\bar{z}^{B}+\frac{1}{4}{\rm{Im}}\mathcal{N}_{IJ}
F^{I}\cdot F^{J}+\frac{1}{4}{\rm{Re}}\mathcal{N}_{IJ}F^{I}\cdot 
\tilde{F}^{J}\;.  \label{action}
\end{equation}

The $n$ complex scalar fields $z^{A}$ of $N=2$ vector multiplets are
coordinates of a special K\"{a}hler manifold. $F^{I}$ are $n+1$ two-forms
representing the gauge field strength two-forms and we have used the
notation $F\cdot F=F_{\mu \nu }F^{^{\mu \nu }}.$ 

A special K\"{a}hler manifold is a K\"{a}hler-Hodge manifold with conditions
on the curvature

\begin{equation}
R_{A\bar{B}C\bar{D}}=g_{A\bar{B}}g_{C\bar{D}}+g_{A\bar{D}}
g_{C\bar{B}}-C_{ACE}C_{\bar{B}\bar{D}\bar{L}}g^{E\bar{L}}.  \label{curv}
\end{equation}
Here $g_{A\bar{B}}=\partial _{A}\partial _{\bar{B}}K$ is the K\"{a}hler
metric, $K$ is the K\"{a}hler potential and $C_{ABC}$ is a completely
symmetric covariantly holomorphic tensor. A K\"{a}hler-Hodge manifold has a 
$U(1)$ bundle whose first Chern class coincides with the K\"{a}hler class,
thus locally the $U(1)$ connection $A$ can be written as

\begin{equation}
A=-\frac{i}{2}(\partial _{A}Kdz^{A}-\partial _{\bar{A}}Kd\bar{z}^{A}).
\label{ivp}
\end{equation}
A useful definition of a special K\"{a}hler manifold can be given by
introducing a ($2n+2)$-dimensional symplectic bundle over the 
K\"{a}hler-Hodge manifold with the covariantly holomorphic sections

\begin{eqnarray}
V &=&\left( 
\begin{array}{c}
L^{I} \\ 
M_{I}
\end{array}
\right) ,\text{ \ \ \ }I=0,...,n  \notag \\
D_{\bar{A}}V &=&\left( \partial _{\bar{A}}-\frac{1}{2}
\partial _{\bar{A}}K\right) V=0.  \label{symplectic}
\end{eqnarray}
These sections obey the symplectic constraint

\begin{equation}
i\langle V,\bar{V}\rangle =i\left( \bar{L}^{I}M_{I}-L^{I}\bar{M}_{I}\right)
=1.
\end{equation}
One also defines

\begin{equation}
U_{A}=D_{A}V=\left( \partial _{A}+\frac{1}{2}\partial _{A}K\right) V=\left( 
\begin{array}{c}
f_{A}^{I} \\ 
h_{AI}
\end{array}
\right) .  \label{dse}
\end{equation}
In general one can write

\begin{equation}
M_{I}=\mathcal{N}_{IJ}L^{J},\text{ \ \ \ \ \ }h_{AI}=\mathcal{\bar{N}}_{IJ}f_{A}^{J}  \label{st}
\end{equation}
where ${\cal{N}}_{IJ}$ is a symmetric complex matrix. It can be demonstrated
that the constraint (\ref{curv}) can be obtained from the integrability
conditions on the following differential constraints

\begin{eqnarray}
U_{A} &=&D_{A}V,  \notag \\
D_{A}U_{B} &=&iC_{ABC}g^{C\bar{D}}\bar{U}_{\bar{D}},  \notag \\
D_{A}\bar{U}_{\bar{B}} &=&g_{A\bar{B}}\bar{V},  \notag \\
D_{A}\bar{V} &=&0,  \notag \\
\langle V,U_{A}\rangle  &=&0.  \label{cons}
\end{eqnarray}
The K\"{a}hler potential is introduced via the definition of the holomorphic
sections

\begin{eqnarray}
\Omega  &=&e^{-K/2}V=\left( 
\begin{array}{c}
X^{I} \\ 
F_{I}
\end{array}
\right) ,\text{ \ \ \ \ }\partial _{\bar{A}}\Omega =0,  \notag \\
\mathcal{D}_{A}\Omega  &=&\left( \partial _{A}+\partial _{A}K\right) \Omega ,
\text{ \ \ \ \ \ }  \notag \\
F_{I}(z) &=&\mathcal{N}_{IJ}X^{J}(z),\text{ \ \ \ \ }\mathcal{D}_{A}F_{I}(z)=
\mathcal{\bar{N}}_{IJ}\mathcal{D}_{A}X^{I}(z).  \label{nr}
\end{eqnarray}
Using (\ref{symplectic}) we obtain 

\begin{equation}
e^{-K}=i\left( \bar{X}^{I}F_{I}-X^{I}\bar{F}_{I}\right) .  \label{kp}
\end{equation}
Here we list some equations coming from special geometry

\begin{eqnarray}
g_{A\bar{B}} &=&\partial _{A}\partial _{\bar{B}}K=-i\langle U_{A},
\bar{U}_{\bar{B}}\rangle =-2{\rm{Im}}\mathcal{N}_{IJ}f_{A}^{I}\bar{f}_{\bar{B}}^{J},
\label{r} \\
g^{A\bar{B}}f_{A}^{I}\bar{f}_{\bar{B}}^{J} &=&-\frac{1}{2}\left( {\rm{Im}}
\mathcal{N}\right) ^{IJ}-\bar{L}^{I}L^{J},  \label{rr} \\
F_{I}\partial _{\mu }X^{I}-X^{I}\partial _{\mu }F_{I} &=&0.  \label{rrr}
\end{eqnarray}

Recently, Euclidean versions of special geometry have been investigated in
the context of Euclidean supergravity theories \cite{mohaupt3}. 
The Euclidean  theories are found by 
replacing $i$ with $e$ in the corresponding Lorentzian versions of the theories, 
where $e$ has the
properties $e^2=1$ and ${\bar{e}}=-e$.
Thus one has, in the Euclidean theory, the para-complex fields
\begin{equation}
L^{I}={\rm{Re}}L^{I}+e{\rm{Im}}L^{I},\text{ \ \ \ }\bar{L}^{I}={\rm{Re}}
L^{I}-e{\rm{Im}}L^{I}
\end{equation}
and as a result all quantities expressed in terms of $L^{I}$ become
para-complex. One can also introduce the so-called adapted coordinates which
are defined as

\bea
L_{\pm }^{I}={\rm{Re}}L^{I}\pm {\rm{Im}}L^{I}.  \label{adapt}
\eea
and also set
\bea
M_{\pm I} = {\rm{Re}}M_I \pm {\rm{Im}}M_I \ .
\eea
It should be noted that the replacement of $i$ by $e$, was first done in the
context of finding D-instanton solutions in type IIB supergravity \cite{gb}.
This replacement is effectively the replacement of the complex structure by
a para-complex structure. Details on para-complex geometry, para-holomorphic
bundles, para-K\"{a}hler manifolds and affine special para-K\"{a}hler
manifolds can be found in \cite{mohaupt1}. The Killing spinor equations in
the Euclidean $N=2$ supergravity theory were recently obtained in \cite{eusy}
by reducing those of the five-dimensional theory given in \cite{GST}.

The equations of special geometry for either signatures can be considered in
a unified manner by introducing the symbol $i_{\epsilon },$ 
$\bar{\imath}_{\epsilon }=-i_{\epsilon },$ where $i_{\epsilon }^{2}=\epsilon ,$ with 
$\epsilon =-1$ for theories with $(1,3)$ signature and $\epsilon =+1$ for
theories with $\left( 0,4\right) $ signature. Note for the adapted
coordinates one uses the definition given in (\ref{adapt}).

From the above equations one can derive some useful relations which will be
needed in our analysis. Using (\ref{symplectic}) and (\ref{dse}), we write 
\begin{eqnarray}
\partial _{\mu }L^{I} &=&\partial _{\bar{A}}L^{I}\partial _{\mu }\bar{z}^{A}
+\partial _{A}L\partial _{\mu }z^{A} \nonumber \\
&=&\frac{1}{2}\partial _{\bar{A}}KL^{I}\partial _{\mu }\bar{z}^{A}+
\mathcal{D}_{A}L^{I}\partial _{\mu }z^{A}-\frac{1}{2}\partial _{A}KL^{I}\partial _{\mu
}z^{A} 
\end{eqnarray}
which implies the relation 

\begin{equation}
\text{\ }\mathcal{D}_{A}L^{I}\partial _{\mu }z^{\alpha }=\text{\ }\partial
_{\mu }L^{J}-\epsilon i_{\epsilon }L^{J}A_{\mu }  \label{usef2}
\end{equation}
after using

\begin{equation}
A=-\frac{i_{\epsilon }}{2}(\partial _{A}Kdz^{A}-\partial _{\bar{A}}Kd\bar{z}^{A}).
\end{equation}
Moreover from (\ref{kp}) we have

\begin{equation}
\partial _{A}Kdz^{A}=-i_{\epsilon }e^{K}\left( \bar{X}^{I}dF_{I}-\bar{F}_{I}dX^{I}\right) .  \label{k}
\end{equation}
Using (\ref{k}), we obtain from (\ref{ivp}) 
\begin{equation}
A=\epsilon \left( L^{I}d\bar{M}_{I}-M_{I}d\bar{L}^{I}\right) .  \label{vp}
\end{equation}
Also using (\ref{symplectic}), (\ref{dse}) and (\ref{st}), one obtains
\begin{eqnarray}
\partial _{\mu }M_{I} &=&\partial _{\bar{A}}M_{I}\partial _{\mu }\bar{z}^{A}+\partial _{A}M_{I}\partial _{\mu }z^{A} \nonumber \\
&=&\frac{1}{2}M_{I}\partial _{\bar{A}}K\partial _{\mu }\bar{z}^{A}+\mathcal{D}_{A}M_{I}\partial _{\mu }z^{A}-\frac{1}{2}M_{I}\partial _{A}K\partial _{\mu
}z^{A} \nonumber  \\
&=&\epsilon i_{\epsilon }M_{I}A_{\mu }+\mathcal{\bar{N}}_{IJ}\mathcal{D}_{A}L^{I}\partial _{\mu }z^{A} \nonumber  \\
&=&\epsilon i_{\epsilon }M_{I}A_{\mu }+\mathcal{\bar{N}}_{IJ}\left( \partial
_{\mu }L^{J}-\epsilon i_{\epsilon }L^{J}A_{\mu }\right) \ .
\end{eqnarray}
This implies that 
\begin{equation}
\partial _{\mu }M_{I}-2{\rm{Im}}\mathcal{N}_{IJ}L^{J}A_{\mu }=
\mathcal{\bar{N}}_{IJ}\partial _{\mu }L^{J}.  \label{usef1}
\end{equation}

It will be convenient to rewrite a number of these conditions
in terms of adapted co-ordinates, which will be used
for the Euclidean calculation in the following section. In particular, 
the relationship between $M_I$ and $L^I$ given in ({\ref{st}}) is
equivalent to
\bea
\label{usefa}
M_{\pm I} = ({\rm{Re}} {\cal{N}}_{IJ} \pm {\rm{Im}} {\cal{N}}_{IJ}) L^J_\pm \ .
\eea
The condition ({\ref{rr}}) is equivalent to
\bea
\label{usefb}
\big( {\rm{Re}}(g^{A {\bar{B}}} {\cal{D}}_{\bar{B}} {\bar{L}}^I)
\pm  {\rm{Im}}(g^{A {\bar{B}}} {\cal{D}}_{\bar{B}} {\bar{L}}^I) \big)
\big({\rm{Re}} ({\cal{D}}_A L^J)\pm {\rm{Im}} ({\cal{D}}_A L^J) \big)
&=& -{1 \over 2} ({\rm{Im}} {\cal{N}})^{IJ} 
\nonumber \\
&-& L^I_\pm L^J_\mp \ ,
\eea
and ({\ref{rrr}}) is equivalent to
\bea
\label{usefc}
M_{\pm I} d L^I_{\pm} - L^I_\pm dM_{\pm I}=0 \ .
\eea
Also, ({\ref{usef2}}) is equivalent to
\bea
\label{usefd}
\partial_\mu L^I_\pm = \big({\rm{Re}} ({\cal{D}}_AL^I) \pm {\rm{Im}} ({\cal{D}}_A L^I) \big)
\partial_\mu \big({\rm{Re}} z^A \pm {\rm{Im}}z^A \big) \pm A_\mu L^I_\pm
\eea
and ({\ref{usef1}}) is equivalent to
\bea
\label{usefe}
\partial_\mu M_{\pm I} = ({\rm Re} {\cal{N}}_{IJ} \pm {\rm Im} {\cal{N}}_{IJ})
\partial_\mu L^J_\pm +2 A_\mu {\rm Im}{\cal{N}}_{IJ} L^J_\pm \ .
\eea
In addition, note that on contracting ({\ref{r}}) with $g^{A \bar{B}}$, and using ({\ref{rr}}), 
one finds that
\bea
{n \over 2} = g^{A {\bar{B}}} g_{A \bar{B}} = {\rm Im}{\cal{N}}_{IJ}  {\rm Im}{\cal{N}}^{IJ} + 2 {\rm Im} {\cal{N}}_{IJ} {\bar{L}}^I L^J
\eea
and hence
\bea
{\rm Im} {\cal{N}}_{IJ} {\bar{L}}^I L^J = -{1 \over 2} -{n \over 4} \ .
\eea
This implies that
\bea
\label{useff}
{\rm Im} {\cal{N}}_{IJ} L^I_+ L^J_- = -{1 \over 2}-{n \over 4}\ .
\eea

\section{Gravitational Instantons}

In this section we classify the gravitational instanton solutions by solving
the Killing spinor equations for the Euclidean theory \cite{eusy}:
\begin{eqnarray}
\left( \nabla _{\mu }-{\frac{1}{2}}A_{\mu }\Gamma _{5}+\frac{i}{4}\Gamma
\cdot F^{I}\left( {\rm{Im}}L^{J}+\Gamma _{5}{\rm{Re}}L^{J}\right) 
({\rm{Im}}\mathcal{N})_{IJ}\Gamma _{\mu }\right) \varepsilon &=&0  \label{grr} \\
\frac{i}{2}({\rm{Im}}\mathcal{N})_{IJ}\Gamma \cdot F^{J}\left[ {\rm{Im}}
(\mathcal{D}_{\bar{B}}{\bar{L}}^{I}g^{A\bar{B}})+\Gamma _{5}{\rm{Re}}
(\mathcal{D}_{\bar{B}}{\bar{L}}^{I}g^{A\bar{B}})\right] \varepsilon
\nonumber \\
+\Gamma^{\mu }\partial _{\mu }\left[ {\rm{Re}}z^{A}-\Gamma _{5}{\rm{Im}}z^{A}\right]
\varepsilon \, &=&0  \ . \label{gee}
\end{eqnarray}

We remark that if $\varepsilon$ is a Killing spinor satisfying ({\ref{grr}}) and ({\ref{gee}}), then so is
$C* \Gamma_5 \varepsilon$, which is moreover linearly independent of $\varepsilon$ (over $\bC$);
here $C*$ is a charge conjugation operator whose construction is defined in terms of spinorial geometry
techniques in \cite{instantons}.
It follows that the complex space of Killing spinors must be of even dimension, i.e. the supersymmetric solutions must preserve
either 4 or 8 {\it real} supersymmetries. If a solution is maximally supersymmetric, then the gaugino Killing spinor equation
({\ref{gee}}) implies that the scalars $z^A$ are constant. It then follows that the gravitino Killing spinor equation reduces
to the gravitino Killing spinor equation of the minimal theory. As the maximally supersymmetric solutions of the 
minimal theory have already been fully classified in \cite{instantons}, for the remainder of this paper we shall consider
solutions preserving half of the supersymmetry.

In order to proceed with the analysis, we define the \textit{spacetime basis} $\mathbf{e}^{1},\mathbf{e}^{2},\mathbf{e}^{\bar{1}},\mathbf{e}^{\bar{2}}$, with respect to which the spacetime metric
is 
\begin{equation}
ds^{2}=2\left( \mathbf{e}^{1}\mathbf{e}^{\bar{1}}+\mathbf{e}^{2}\mathbf{e}^{\bar{2}}\right) .
\end{equation}
The space of Dirac spinors is taken to be the complexified space of forms on 
$\mathbb{R}^{2}$, with basis $\{1,e_{1},e_{2},e_{12}=e_{1}\wedge e_{2}\}$; a
generic Dirac spinor $\varepsilon $ is a complex linear combination of these
basis elements. In this basis, the action of the Dirac matrices $\Gamma _{m}$
on the Dirac spinors is given by 
\begin{equation}
\Gamma _{m}=\sqrt{2}i_{e_{m}},\qquad \Gamma _{\bar{m}}=\sqrt{2}e_{m}\wedge
\end{equation}
for $m=1,2$. We also define
\begin{equation}
\Gamma _{5}=\Gamma _{1\bar{1}2\bar{2}}
\end{equation}
which acts on spinors via 
\begin{equation}
\Gamma _{5}1=1,\qquad \Gamma _{5}e_{12}=e_{12},\qquad \Gamma
_{5}e_{m}=-e_{m}\quad m=1,2.
\end{equation}
With this representation of the Dirac matrices acting on spinors,
the resulting linear system obtained from ({\ref{grr}}) and ({\ref{gee}}) is 
listed in Appendix A.

There are three non-trivial orbits of $Spin(4)=Sp(1)\times Sp(1)$ acting on the space of Dirac spinors. In our
notation, one can use $SU(2)$ transformations to rotate a generic spinor $\epsilon $ into the canonical form \cite{instantons, bryant}
\begin{equation}
\varepsilon =\lambda 1+\sigma e_{1},  \label{canon}
\end{equation}
where $\lambda ,\sigma \in \mathbb{R}$. The three orbits mentioned above
correspond to the cases $\lambda =0$, $\sigma \neq 0$; $\lambda \neq 0$, $\sigma =0$ and $\lambda \neq 0$, $\sigma \neq 0$. The orbits corresponding
to $\lambda =0$, $\sigma \neq 0$ and $\lambda \neq 0$, $\sigma =0$ are
equivalent under the action of $Pin(4)$. We shall treat these orbits separately.

\subsection{Solutions with $\lambda \neq 0$ and $\sigma \neq 0$}

For solutions with $\lambda \neq 0$ and $\sigma \neq 0$, the analysis
of the linear system (\ref{linearsystem}) obtained from the gravitino equation produces
the following geometric conditions
\begin{eqnarray}
\omega _{1,2\bar{2}} &=&\partial _{1}\log \frac{\sigma }{\lambda }+A_{1},
\text{ \ \ \ \ \ \ \ \ \ \ \ \ \ }\omega _{2,1\bar{1}}=\partial _{2}\log 
\frac{\lambda }{\sigma }-A_{2},\text{\ \ }  \notag \\
\omega _{1,1\bar{1}} &=&-\partial _{1}\log \lambda \sigma ,\text{ \ \ \ \ \
\ \ \ \ \ \ \ \ \ \ \ \ \ \ }\omega _{2,2\bar{2}}=\partial _{2}\log \lambda
\sigma ,  \notag \\
\omega _{1,21} &=&2\partial _{2}\log \lambda -A_{2},\text{ \ \ \ \ \ \ \ \ \
\ \ \ \ }\omega _{\bar{1},2\bar{1}}=2\partial _{2}\log \sigma +A_{2},  \notag
\\
\omega _{2,\bar{2}1} &=&-2\partial _{1}\log \sigma -A_{1},\text{ \ \ \ \ \ \
\ \ \ \ \ }\omega _{\bar{2},21}=-2\partial _{1}\log \lambda +A_{1},\text{ \ }
\notag \\
\omega _{\bar{1},21} &=&\omega _{2,21}=\omega _{1,2\bar{1}}=
\omega _{2,2\bar{1}}=0\text{,}  \label{gcc}
\end{eqnarray}
as well as the following conditions involving the gauge field strengths
\begin{eqnarray}
({\rm{Im}}\,\mathcal{N})_{IJ}\left( F_{2\bar{2}}^{I}+F_{1\bar{1}}^{I}\right)
L_{+}^{J} &=&-\frac{\sqrt{2}i}{\lambda \sigma }\left( \partial
_{1}-A_{1}\right) \lambda ^{2},  \notag \\
({\rm{Im}}\,\mathcal{N})_{IJ}\left( F_{2\bar{2}}^{I}-F_{1\bar{1}}^{I}\right)
L_{-}^{J} &=&\frac{\sqrt{2}i}{\lambda \sigma }\left( \partial _{\bar{1}}+
A_{\bar{1}}\right) \sigma ^{2},  \notag \\
({\rm{Im}}\,\mathcal{N})_{IJ}F_{21}^{I}L_+^{J} &=&\frac{i}{\sqrt{2}\lambda
\sigma }\left( \partial _{2}-A_{2}\right) \lambda ^{2},  \notag \\
({\rm{Im}}\,\mathcal{N})_{IJ}F_{2\bar{1}}^{I}L_{-}^{J} &=&
-\frac{i}{\sqrt{2}\lambda \sigma }\left( \partial _{2}+A_{2}\right) \sigma ^{2}.
\label{graveq}
\end{eqnarray}

The geometric constraints \ (\ref{gcc}) imply the following
\begin{eqnarray}
d\mathbf{e}^{1} &=&-\partial _{1}\log \lambda \sigma \mathbf{e}^{1}\wedge 
\mathbf{e}^{\bar{1}}+\left( 2\partial _{2}\log \sigma +A_{2}\right) \mathbf{e}^{\bar{1}}\wedge \mathbf{e}^{2}+\left( 2\partial _{\bar{2}}\log \lambda -A_{\bar{2}}\right) \mathbf{e}^{\bar{1}}\wedge \mathbf{e}^{\bar{2}}  \notag \\
&&+\left( \partial _{2}\log \frac{\sigma }{\lambda }+A_{2}\right) \mathbf{e}^{1}\wedge \mathbf{e}^{2}  \notag \\
&&+\left( \partial _{\bar{2}}\log \frac{\lambda }{\sigma }-A_{\bar{2}}\right) \mathbf{e}^{1}\wedge \mathbf{e}^{\bar{2}}+2\left( \partial _{1}\log 
\frac{\sigma }{\lambda }+A_{1}\right) \mathbf{e}^{2}\wedge \mathbf{e}^{\bar{2}}  \notag \\
d\mathbf{e}^{2} &=&-d\left( \log \lambda \sigma \right) \wedge \mathbf{e}^{2}
\end{eqnarray}
implying that

\begin{equation}
d\left( \lambda \sigma \left( \mathbf{e}^{1}+\mathbf{e}^{\bar{1}}\right)
\right) =0.
\end{equation}
Thus we introduce three real local coordinates $x,y$ and $z,$ such that
\begin{equation}
\left( \mathbf{e}^{1}+\mathbf{e}^{\bar{1}}\right) =\frac{\sqrt{2}}{\lambda
\sigma }dx,\text{ \ \ \ \ }\mathbf{e}^{2}=\frac{1}{\sqrt{2}\lambda \sigma }
\left( dy+idz\right) \ .
\end{equation}
Furthermore, the vector defined by 
\begin{equation}
V=i\lambda \sigma \left( \mathbf{e}^{1}-\mathbf{e}^{{\bar{1}}}\right)
\end{equation}
is a Killing vector, and so we introduce a local-coordinate $\tau$ such that
\bea
V= \sqrt{2} {\partial \over \partial \tau}
\eea
and 
\begin{equation}
\sqrt{2}\left( \mathbf{e}^{1}-\mathbf{e}^{\bar{1}}\right) =-2i\lambda \sigma
(d\tau +\phi )
\end{equation}
where $\phi = \phi_x dx+ \phi_y dy + \phi_z dz$ is a 1-form. 
We remark that the conditions imposed on the geometry by the gravitino Killing spinor equations imply that
\bea
\label{acon1}
\bigg( A+d \left( \log \frac{\sigma }{\lambda }\right) \bigg)_\tau =0 \ .
\eea

So, in the co-ordinates $\tau, x, y, z$, the metric is
\begin{equation}
ds^{2}=\left( \lambda \sigma \right) ^{2}(d\tau +\phi )^{2}+\frac{1}{\left(
\lambda \sigma \right) ^{2}}\left( dx^{2}+dy^{2}+dz^{2}\right)
\end{equation}
where $\lambda \sigma$ and $\phi$ are independent of $\tau$, and
$\phi$ satisfies
\begin{equation}
d\phi =\frac{2}{\left( \lambda \sigma \right) ^{2}}\hat{\ast}\left[ A+d \left( \log \frac{\sigma }{\lambda }\right) \right] .  \label{dphi}
\end{equation}
where $\hat{\ast}$ denotes the Hodge dual on  $\bR^3$ equipped with metric $dx^2+dy^2+dz^2$
and volume form $dx \wedge dy \wedge dz$. 

Next consider we consider the linear system (\ref{gaconditions}) derived from (\ref{gee}).
First, one finds that
\bea
{\cal{L}}_V z^A =0 \ .
\eea
This condition implies that
\bea
A_\tau=0
\eea
and hence the $\tau$-independence of $\lambda \sigma$, together with ({\ref{acon1}}), imply that both
$\lambda$ and $\sigma$ are independent of $\tau$.

Next, using ({\ref{usefb}}), together with ({\ref{graveq}}), one obtains
\begin{eqnarray}
iF_{21}^{I}-L_{-}^{I}\frac{\sqrt{2}}{\lambda \sigma }\left( \partial
_{2}-A_{2}\right) \lambda ^{2}-\sqrt{2}\frac{\sigma }{\lambda }\left(
\partial _{2}-A_{2}\right) L_{+}^{I} &=&0,  \notag \\
iF_{2\bar{1}}^{I}+L_{+}^{I}\frac{\sqrt{2}}{\lambda \sigma }\left( \partial
_{2}+A_{2}\right) \sigma ^{2}+\sqrt{2}\frac{\lambda }{\sigma }\left(
\partial _{2}+A_{2}\right) L_{-}^{I} &=&0,  \notag \\
-\frac{i}{2}\left( F_{2\bar{2}}^{I}-F_{1\bar{1}}^{I}\right) +L_{+}^{I}\frac{\sqrt{2}}{\lambda \sigma }\left( \partial _{\bar{1}}+A_{\bar{1}}\right)
\sigma ^{2}+\sqrt{2}\frac{\lambda }{\sigma }\left( \partial
_{1}+A_{1}\right) L_{-}^{I} &=&0,  \notag \\
\frac{i}{2}\left( F_{2\bar{2}}^{I}+F_{1\bar{1}}^{I}\right) +L_{-}^{I}\frac{\sqrt{2}}{\lambda \sigma }\left( \partial _{1}-A_{1}\right) \lambda ^{2}+
\sqrt{2}\frac{\sigma }{\lambda }\left( \partial _{1}-A_{1}\right) L_{+}^{I}
&=&0,
\end{eqnarray}
from which we obtain
\begin{eqnarray}
F_{1\bar{1}}^{I} &=&i\partial _{x}\left( \sigma ^{2}L_{+}^{I}+\lambda
^{2}L_{-}^{I}\right) , \nonumber \\
F_{2\bar{2}}^{I} &=&i\partial _{x}\left( \lambda ^{2}L_{-}^{I}-\sigma
^{2}L_{+}^{I}\right) +2i\sigma ^{2}\left( \partial _{x}-A_{x}\right)
L_{+}^{I}-2i\lambda ^{2}\left( \partial _{x}+A_{x}\right) L_{-}^{I} \ , \nonumber \\
F_{2\bar{1}}^{I} &=&iL_{+}^{I}\left( \left( \partial _{y}-i\partial
_{z}\right) \sigma ^{2}+\sigma ^{2}\left( A_{y}-iA_{z}\right) \right)
\nonumber \\
&+&i\left( \lambda ^{2}\left( \partial _{y}-i\partial _{z}\right) +\lambda
^{2}\left( A_{y}-iA_{z}\right) \right) L_{-}^{I} \ , \nonumber \\
F_{21}^{I} &=&iL_{-}^{I}\left( -\left( \partial _{y}-i\partial _{z}\right)
\lambda ^{2}+\lambda ^{2}\left( A_{y}-iA_{z}\right) \right) 
\nonumber \\
&-&i\left( \sigma
^{2}\left( \partial _{y}-i\partial _{z}\right) -\sigma ^{2}\left(
A_{y}-iA_{z}\right) \right) L_{+}^{I}.
\nonumber \\
\end{eqnarray}
In terms of the local co-ordinates $\tau, x, y, z$, the gauge field strengths are
\begin{equation}
F^{I}=-d \left[ \left( \sigma ^{2}L^{I}+\lambda ^{2}L_{-}^{I}\right)
(d\tau +\phi )\right] +\hat{\ast} d \left[ \frac{L_{+}^{I}}{\lambda ^{2}}-\frac{L_{-}^{I}}{\sigma ^{2}}\right] \ .
\end{equation}
Thus the Bianchi identity implies that
\begin{equation}
\hat{\nabla}^{2}\left[ \frac{L_{+}^{I}}{\lambda ^{2}}-
\frac{L_{-}^{I}}{\sigma ^{2}}\right] =0.
\end{equation}
where ${\hat{\nabla}}^2$ is the Laplacian on $\bR^3$.
The dual gauge field strength ${\tilde{F}}^I$ is given by 
\bea
{\tilde{F}}^I_{\mu_1 \mu_2} = {1 \over 2} \epsilon_{\mu_1 \mu_2}{}^{\nu_1 \nu_2} F^I_{\nu_1 \nu_2}
\eea
where the volume form satisfies $\epsilon_{1 {\bar{1}} 2 {\bar{2}}} =1$. Hence
\begin{equation}
\tilde{F}^I_{2\bar{2}}=-F^I_{1\bar{1}},\text{ \ \ \ \ \ }\tilde{F}^I_{1\bar{1}}=
-F^I_{2\bar{2}},\text{ \ \ \ }\tilde{F}^I_{1\bar{2}}=F^I_{1\bar{2}},\text{ \ \ \ \ \ }
\tilde{F}^I_{12}=-F^I_{12}.
\end{equation}
In terms of the local co-ordinates $\tau, x, y, z$ one finds
\bea
{\tilde{F}}^I &=& \bigg( L^I_- d \lambda^2 - L^I_+ d \sigma^2 - \lambda^2 d L^I_-
+ \sigma^2 d L^^I_+ - 2 (\lambda^2 L^I_- + \sigma^2 L^I_+)A \bigg) \wedge \big( d \tau + \phi \big)
\nonumber \\
&-&{1 \over \lambda^2 \sigma^2} {\hat{\ast}} d \big( \lambda^2 L^I_- + \sigma^2 L^I_+ \big) \ .
\eea

Evaluating ${\rm{Re}}\mathcal{N}_{IJ}F^{J}+{\rm{Im}}\mathcal{N}_{IJ}
{\tilde{F}}^{J}$ and making use of (\ref{usefe}) we obtain

\begin{equation}
{\rm{Re}}\mathcal{N}_{IJ}F^{J}+{\rm{Im}}\mathcal{N}_{IJ} {\tilde{F}}^{J}=
-d\left[ \left[ \sigma ^{2}M_{+I}+\lambda ^{2}M_{-I}\right] (d\tau +\phi )\right]
+{\hat{\ast}} d\left[ \frac{M_{+I}}{\lambda ^{2}}-\frac{M_{-I}}{\sigma ^{2}}\right] .
\end{equation}
The gauge field equations are 
\begin{equation}
d\left[ {\rm{Re}}\mathcal{N}_{IJ}F^{J}+{\rm{Im}}\mathcal{N}_{IJ}{\tilde{F}}^{J}\right] =0,
\end{equation}
which implies that
\begin{equation}
{\hat{\nabla}}^{2}\left[ \frac{M_{+I}}{\lambda ^{2}}-\frac{{M}_{-I}}{\sigma ^{2}}\right] =0.
\end{equation}
Therefore Bianchi identities and Maxwell's equations imply

\begin{equation}
\frac{M_{+I}}{\lambda ^{2}}-\frac{M_{-I}}{\sigma ^{2}}=H_{I},\text{ \ \ \ \
\ }\frac{L_{+}^{I}}{\lambda ^{2}}-\frac{L_{-}^{I}}{\sigma ^{2}}=H^{I}.
\label{ste}
\end{equation}
These are the Euclidean version of the stabilisation conditions. Also (\ref{ste}) implies

\begin{equation}
\frac{1}{\sigma ^{2}}=L_+^{I}H_{I}-M_{+I}H^{I},\text{ \ \ \ \ \ \ \ }\frac{1}{\lambda ^{2}}=L_{-}^{I}H_{I}-M_{-I}H^{I}.
\end{equation}
From the stablisation conditions (\ref{ste}) and (\ref{vp}) we obtain

\begin{equation}
A=\frac{\lambda ^{2}\sigma ^{2}}{2}\left( H_{I}dH^{I}-H^{I}dH_{I}\right)
+d\log \frac{\lambda }{\sigma }.  \label{pr}
\end{equation}
Returning to (\ref{dphi}), (\ref{pr}) implies that
\begin{equation}
d\phi = {\hat{\ast}} \left[ \left( H_{I}dH^{I}-H^{I}dH_{I}\right) \right] .
\end{equation}
This system of equations for Euclidean instantons can be analysed in a similar way to the analysis
performed for the corresponding black hole solutions in the Lorentzian theory \cite{denef1, denef2}.

\subsection{Solutions with $\protect\sigma =0,$ $\protect\lambda \neq 0$ and $\protect\lambda =0,$ $\protect\sigma \neq 0$}

The analysis of the linear system in (\ref{linearsystem}) 
for the case of $\sigma =0,$ $\lambda \neq 0,$ 
implies that the spin connections satisfy
\begin{equation}
\omega _{\mu ,12}=0,\text{ \ \ \ \ \ \ \ }\omega _{\mu ,1\bar{1}}+\omega
_{\mu ,2\bar{2}}=0.\text{\ }  \label{sc}
\end{equation}
The conditions on the spin connection (\ref{sc}) imply that the (anti-self-dual) almost
complex structures\textbf{\ }$\mathbf{I}^{1},$\textbf{\ }$\mathbf{I}^{2},$\textbf{\ }$\mathbf{I}^{3}$ defined by
\begin{equation}
\mathbf{I}^{1}=\left( \mathbf{e}^{12}+\mathbf{e}^{\bar{1}\bar{2}}\right) ,
\text{ \ \ \ \ }\mathbf{I}^{2}=i\left( \mathbf{e}^{12}-\mathbf{e}^{\bar{1}\bar{2}}\right) ,\text{ \ \ \ \ }\mathbf{I}^{3}=i\left( \mathbf{e}^{1\bar{1}}+\mathbf{e}^{2\bar{2}}\right) 
\end{equation}
and which satisfy the algebra of the imaginary unit quaternions, are
covariantly constant with respect to the Levi-Civita connection, i.e. the
manifold is hyper-K\"{a}hler. 
The remaining conditions from ({\ref{linearsystem}}) are
\bea
\label{aac}
A =2 d \log \lambda
\eea
and
\bea
{\rm Im} {\cal{N}}_{IJ} (F^I_{1 \bar{1}}- F^I_{2 \bar{2}}) L^J_- &=&0 \ ,
\nonumber \\
{\rm Im} {\cal{N}}_{IJ} F^I_{1 {\bar{2}}} L^J_- &=&0 \ .
\eea
It is also straightforward to show that ({\ref{gaconditions}}) implies that
\bea
\label{gsd1}
F^N_{1 \bar{1}} + F^N_{2 \bar{2}} &=& -2 {\rm{Im}} {\cal{N}}_{IJ} L^I_+ (F^J_{1 \bar{1}} + F^J_{2 \bar{2}}) L^N_- \ .
\nonumber \\
F^N_{12} &=& -2 {\rm Im} {\cal{N}}_{IJ} L^I_+ F^J_{12} L^N_- \ .
\eea
In the non-minimal theory, ({\ref{gsd1}}) implies that $F^I$ are self-dual.
To see this, consider first $F^I_{1 \bar{1}}+F^I_{2 \bar{2}}$; the first condition in ({\ref{gsd1}}) implies that
\bea
F^I_{1 \bar{1}} + F^I_{2 \bar{2}} = h L^I_-
\eea
where
\bea
h = -2 {\rm Im} {\cal{N}}_{IJ} L^I_+ (F^J_{1 \bar{1}} + F^J_{2 \bar{2}})
\eea
and hence
\bea
h = -2h {\rm Im}{\cal{N}}_{IJ} L^I_+ L^J_- = -2h (-{1 \over 2}-{n \over 4})
\eea
where we have made use of ({\ref{useff}}).
So, for the case of the non-minimal theory, $n\geq 1$, and this condition implies that $h=0$, and hence
\bea
F^I_{1 \bar{1}}+F^I_{2 \bar{2}} =0 \ .
\eea
Similar reasoning applied to the second condition in ({\ref{gsd1}}) also implies that $F^I_{12}=0$.
It follows that $F^I. ~ {\bf{I}}^1= F^I . ~ {\bf{I}}^2 = F^I . ~ {\bf{I}}^3=0$, so as the hypercomplex structures are anti-self-dual,
the $F^I$ must be self-dual.

Next, consider the Einstein field equations; as the manifold is hyper-K\"ahler there is no contribution from the curvature terms.
Also, as the $F^I$ are self-dual, the contribution from the gauge field strengths also vanishes.
So, on taking the trace of the Einstein equations, one obtains
\bea
g_{A \bar{B}} \partial_\mu z^A \partial^\mu z^{\bar{B}} =0 \ .
\eea
Assuming that the scalar manifold metric is positive definite, this implies that the scalars are constant,
and ({\ref{aac}}) then implies that $\lambda$ is constant as well.

The analysis of the case $\lambda =0$, $\sigma \neq 0$ proceeds in exactly the same fashion. 
In particular, the conditions on the geometry, and on the gauge field strengths, are identical
to those of the $\lambda \neq 0$, $\sigma=0$ case, modulo the interchange $L^I_+ \leftrightarrow L^I_-$
and (space-time frame index) $1 \leftrightarrow {\bar{1}}$ throughout. Hence, it follows that the manifold is again hyper-K\"ahler, though now with
self-dual hyper-complex structures, and the gauge field strengths must (for the non-minimal theory) be anti-self-dual.
The scalars are again constant, as is $\sigma$.

\section{Summary}

In this paper we have classified the instanton solutions admitting Killing
spinors for Euclidean $N=2$ supergravity theory coupled to vector
multiplets. The half-supersymmetric solutions are either hyper-K\"ahler with
constant scalars and self-dual (or anti-self-dual) gauge field strengths, or they are
Euclidean analogues of the black hole solutions found in 
\cite{BLS}. The stationary black holes of \cite{BLS} were shown to be the
unique solutions with time-like Killing vector, and admitting half of
supersymmetry, in the systematic analysis of \cite{ortin}. They can also be
obtained using spinorial geometry techniques. Employing the results of \cite{4spinor},  the time-like Killing spinors can be written
in the canonical form
\begin{equation}
\varepsilon =1+\beta e^{2}
\end{equation}
which is obtained by gauge fixing the generic spinor of the form
\begin{equation}
\varepsilon =\lambda 1+\mu ^{i}e^{i}+\sigma e^{12}
\end{equation}
where $e^{1}$, $e^{2}$ are 1-forms on $\mathbb{R}^{2}$, and $i=1,2$; $e^{12}=e^{1}\wedge e^{2}$. $\lambda $, $\mu ^{i}$ and $\sigma $ are complex
functions. The analysis of the Killing spinor equations then gives the
solutions

\begin{eqnarray}
ds^{2} &=&-|\beta |^{2}(dt+\sigma )^{2}+\frac{1}{|\beta |^{2}}\left( \left(
dx\right) ^{2}+\left( dy\right) ^{2}+\left( dz\right) ^{2}\right)  \notag \\
d\sigma &=&-\ast _{3}\left[ \left( \left( H_{I}dH^{I}-H^{I}dH_{I}\right)
\right) \right]
\end{eqnarray}
where $\beta $ is a complex $t$-independent function. The gauge
field strengths and scalars are given by

\begin{eqnarray}
F^{I} &=&-\hat{d}\left[ (\frac{L^{I}}{\bar{\beta}}+\frac{\bar{L}^{I}}{\beta })(dt+\sigma )\right] -\ast _{3}\hat{d}H^{I}
\end{eqnarray}
and
\begin{eqnarray}
i\left( \frac{L^{I}}{\bar{\beta}}-\frac{\bar{L}^{I}}{\beta }\right)  &=&H^{I}
\text{, \ \ \ \ \ \ \ \ \ }i\left( \frac{M_{I}}{\bar{\beta}}-\frac{\bar{M}_{I}}{\beta }\right) =H_{I},  \label{stab}
\end{eqnarray}
where $H^{I}$ and $H_{I}$ are harmonic functions. The equations (\ref{stab})
are the so-called generalised stabilisation equations for the scalar fields. 

Recently, instanton solutions of Einstein-Maxwell theory with non-zero
cosmological constant were considered in \cite{firstpaper}. In the analysis
of the particular case with anti-self dual Maxwell field, the field
equations of supersymmetric solutions were shown to reduce to the
Einstein--Weyl system in three dimensions \cite{Hi} which is integrable by a
twistor construction. Also, it was demonstrated that the Maxwell field
anti-self-duality implies Weyl tensor  anti-self-duality.
Moreover, interesting relations were discovered between gravitational
instantons and the $SU(\infty )$ Toda equation. Following on from this, the
anti-self-duality condition on the Maxwell field was relaxed, and supersymmetric
gravitational instanton solutions were classified using spinorial geometry
techniques \cite{mtjs}. An important generalisation of our work is the
construction of Euclidean gauged supergravity theories and the analysis of
their gravitational instanton solutions and their relations to Toda theories
and integrable models. Lifting the solutions of this paper to higher
dimensions as well as the analysis of the instanton moduli spaces are also
left for future investigation.

\newpage

 \setcounter{section}{0}

\appendix{Linear Systems}

The (non-vanishing) actions of the Dirac matrices on the spinors are
given by

\begin{eqnarray}
\text{\ }\Gamma _{1}e_{1} &=&\Gamma _{2}e_{2}=\sqrt{2}1,\text{ \ \ \ \ \ \ \
\ \ \ \ \ \ }  \notag \\
\Gamma _{_{\bar{1}}}1 &=&-\Gamma _{2}e_{12}=\sqrt{2}e_{1},\text{ \ \ } 
\notag \\
\Gamma _{_{\bar{1}}}e_{2} &=&-\Gamma _{\bar{2}}e_{1}=\sqrt{2}e_{12},\text{ }
\notag \\
\Gamma _{\bar{2}}1 &=&\Gamma _{1}e_{12}=\sqrt{2}e_{2}\text{,\ \ \ \ }
\end{eqnarray}
and one also obtains, for a 2-form $T$,
\begin{eqnarray}
T^{ab}\Gamma _{ab}1 &=&2\left( T^{2\bar{2}}+T^{1\bar{1}}\right)1 -4T^{\bar{2}\bar{1}}e_{12} \ , \notag \\
T^{ab}\Gamma _{ab}e_{1} &=&2\left( T^{2\bar{2}}-T^{1\bar{1}}\right)
e_{1}+4T^{\bar{2}1}e_{2} \ , \notag \\
T^{ab}\Gamma _{ab}e_{2} &=&-2\left( T^{2\bar{2}}-T^{1\bar{1}}\right)
e_{2}-4T^{2\bar{1}}e_{1} \ , \notag \\
T^{ab}\Gamma _{ab}e_{12} &=&-2\left( T^{2\bar{2}}+T^{1\bar{1}}\right)
e_{12}+4T^{21}1 \ .
\end{eqnarray}

The linear system obtained from (\ref{grr}) is
\begin{eqnarray}
\partial _{1}\lambda -\frac{\lambda }{2}\left( \omega _{1,2\bar{2}}+\omega
_{1,1\bar{1}}\right) -\frac{i\sigma }{\sqrt{2}}({\rm{Im}}\,\mathcal{N})_{IJ}L_{+}^{J}\left( F_{2\bar{2}}^{I}+F_{1\bar{1}}^{I}\right) -\frac{\lambda }{2}A_{1} &=&0,  \notag \\
\partial _{1}\sigma -\frac{\sigma }{2}\left( \omega _{1,2\bar{2}}-\omega
_{1,1\bar{1}}\right) +\frac{\sigma }{2}A_{1} &=&0,  \notag \\
\partial _{\bar{1}}\lambda -\frac{\lambda }{2}\left( \omega _{\bar{1},2\bar{2}}+\omega _{\bar{1},1\bar{1}}\right) -\frac{\lambda }{2}A_{\bar{1}} &=&0, 
\notag \\
\partial _{\bar{1}}\sigma -\frac{\sigma }{2}\left( \omega _{\bar{1},2\bar{2}}-\omega _{\bar{1},1\bar{1}}-A_{\bar{1}}\right) +\frac{i\lambda }{\sqrt{2}}({\rm{Im}}\,\mathcal{N})_{IJ}\left( F_{2\bar{2}}^{I}-F_{1\bar{1}}^{I}\right)
L_{-}^{J} &=&0,  \notag \\
\partial _{2}\lambda -\frac{\lambda }{2}\left( \omega _{2,2\bar{2}}+\omega
_{2,1\bar{1}}+A_{2}\right)  &=&0,  \notag \\
\partial _{2}\sigma -\frac{\sigma }{2}\left( \omega _{2,2\bar{2}}-\omega
_{2,1\bar{1}}-A_{2}\right)  &=&0,  \notag \\
\partial _{\bar{2}}\lambda -\frac{\lambda }{2}\left( \omega _{\bar{2},2\bar{2}}+\omega _{\bar{2},1\bar{1}}\right) -i\sigma \sqrt{2}({\rm{Im}}\,\mathcal{N})_{IJ}F_{\bar{2}\bar{1}}^{I}L_+^{J}-\frac{\lambda }{2}A_{\bar{2}} &=&0,  \notag
\\
\partial _{\bar{2}}\sigma -\frac{\sigma }{2}\left( \omega _{\bar{2},2\bar{2}}-\omega _{\bar{2},1\bar{1}}\right) +\sqrt{2}i\lambda ({\rm{Im}}\,\mathcal{N})_{IJ}L_{-}^{J}F_{\bar{2}1}^{I}+\frac{\sigma }{2}A_{\bar{2}} &=&0,  \notag \\
\lambda \omega _{\bar{1},21} &=&0,  \notag \\
\lambda \omega _{2,21} &=&0,  \notag \\
\sigma \omega _{1,2\bar{1}} &=&0,  \notag \\
\sigma \omega _{2,2\bar{1}} &=&0,  \notag \\
\sigma \omega _{\bar{1},2\bar{1}}-\sqrt{2}i\lambda ({\rm{Im}}\,\mathcal{N})_{IJ}F_{2\bar{1}}^{I}L_{-}^{J} &=&0,  \notag \\
\sigma \omega _{\bar{2},2\bar{1}}-\frac{i\lambda }{\sqrt{2}}({\rm{Im}}\,\mathcal{N})_{IJ}\left( F_{2\bar{2}}^{I}-F_{1\bar{1}}^{I}\right) L_{-}^{J}
&=&0,  \notag \\
\lambda \omega _{1,21}+\sqrt{2}i\sigma ({\rm{Im}}\,\mathcal{N})_{IJ}F_{21}^{I}L_{+}^{J} &=&0,  \notag \\
\lambda \omega _{\bar{2},21}+\frac{\sigma }{\sqrt{2}}i({\rm{Im}}\,\mathcal{N})_{IJ}\left( F_{2\bar{2}}^{I}+F_{1\bar{1}}^{I}\right) L_{+}^{J} &=&0.
\label{linearsystem}
\end{eqnarray}
The linear system obtained from (\ref{gee}) is
\begin{eqnarray}
-i \lambda {\rm Im}{\cal{N}}_{IJ} (F^J_{1 \bar{1}} + F^J_{2 \bar{2}})
\bigg({\rm Im}({\cal{D}}_{\bar{B}}{\bar{L}}^I g^{A \bar{B}})
+ {\rm Re}({\cal{D}}_{\bar{B}}{\bar{L}}^I g^{A \bar{B}})\bigg)
\nonumber \\
\qquad \qquad+ \sqrt{2} \sigma \partial_{\bar{1}} \bigg( {\rm Re} z^A + {\rm Im} z^A \bigg)&=&0 \ ,
\nonumber \\
i \sigma {\rm Im}{\cal{N}}_{IJ} (F^J_{1 \bar{1}} - F^J_{2 \bar{2}})
\bigg({\rm Im}({\cal{D}}_{\bar{B}}{\bar{L}}^I g^{A \bar{B}})
- {\rm Re}({\cal{D}}_{\bar{B}}{\bar{L}}^I g^{A \bar{B}})\bigg)
\nonumber \\
\qquad \qquad+ \sqrt{2} \lambda \partial_1 \bigg( {\rm Re} z^A - {\rm Im} z^A \bigg)&=&0 \ ,
\nonumber \\
2i \sigma {\rm Im}{\cal{N}}_{IJ} F^J_{2 \bar{1}}
\bigg({\rm Im}({\cal{D}}_{\bar{B}}{\bar{L}}^I g^{A \bar{B}})
- {\rm Re}({\cal{D}}_{\bar{B}}{\bar{L}}^I g^{A \bar{B}})\bigg)
+ \sqrt{2} \lambda \partial_2 \bigg( {\rm Re} z^A - {\rm Im} z^A \bigg) &=&0 \ ,
\nonumber \\
2i \lambda {\rm Im}{\cal{N}}_{IJ} F^J_{2 \bar{1}}
\bigg({\rm Im}({\cal{D}}_{\bar{B}}{\bar{L}}^I g^{A \bar{B}})
+ {\rm Re}({\cal{D}}_{\bar{B}}{\bar{L}}^I g^{A \bar{B}})\bigg)
- \sqrt{2} \sigma \partial_2 \bigg( {\rm Re} z^A + {\rm Im} z^A \bigg) &=&0 \ .
\nonumber \\
  \label{gaconditions}
\end{eqnarray}

\vskip 0.5cm
\noindent{\bf Acknowledgements} \vskip 0.1cm
\noindent  JG is supported by the STFC grant, ST/1004874/1.

\end{document}